\documentclass[aps,pra,twocolumn,groupedaddress,floatfix]{revtex4-1}

\usepackage{graphicx}
\usepackage{dcolumn}
\usepackage{bm}
\usepackage{amsmath}
\usepackage{xcolor}
\usepackage{braket}

\usepackage{bm}
\usepackage[utf8]{inputenc}
\usepackage[T1]{fontenc}

\usepackage{hyperref}
\hypersetup{
breaklinks=true,
    colorlinks=true,
    linkcolor=blue,
    citecolor=blue,
    filecolor=magenta,
    urlcolor=cyan,
}

\usepackage{amsfonts}
\usepackage{blkarray}
\usepackage{amssymb}
\usepackage{multirow}
\usepackage{mathrsfs}

\begin{document}

\title{Excitation spectra of two correlated electrons in a quantum dot}

\author{T. Ihn, C. Ellenberger, K. Ensslin}
\affiliation{Solid State Physics Laboratory, ETH Zurich, CH-8093 Zurich, Switzerland}
\author{Constantine Yannouleas, Uzi Landman}
\email{Constantine.Yannouleas@physics.gatech.edu}
\email{Uzi.Landman@physics.gatech.edu}
\affiliation{School of Physics, Georgia Institute of Technology,
             Atlanta, Georgia 30332-0430}
\author{D.C. Driscoll, A.C. Gossard}
\affiliation{Materials Department, University of California, Santa Barbara, CA 93106}

\begin{abstract}
Measurements and a theoretical interpretation of the excitation spectrum of a two-electron 
quantum dot fabricated on a parabolic Ga[Al]As quantum well are reported. Experimentally, excited states 
are found beyond the well-known lowest singlet- and triplet states. These states can be reproduced in an 
exact diagonalization calculation of a parabolic dot with moderate in-plane anisotropy. The calculated 
spectra are in reasonable quantitative agreement with the measurement, and suggest that correlations 
between the electrons play a significant role in this system. Comparison of the exact results with the
restricted Hartree-Fock and the generalized Heitler-London approach shows that the latter is more 
appropriate for this system because it can account for the spatial correlation of the electron states.
\end{abstract}

\maketitle

Two-electron quantum dots (2eQDs) are the simplest man-made 
structures that allow to study the effects
of electron-electron interaction including exchange and 
correlation. Pioneering experiments measuring the
excitation spectrum of such systems with finite bias spectroscopy
have been reported on vertical few-electron
quantum dots \cite{schm95,kouw97}. Measurements on lateral dots became
possible \cite{kyri02,zumb04,yann06} driven by the potential use of such
systems for the implementation of qubits \cite{loss98}.
Here we present finite bias spectroscopy measurements
performed on a system designed to have a $g$-factor
close to zero and a small lateral anisotropy. We find
that the magnetic field dependent excited state spectrum
found experimentally can be closely reproduced by exact
diagonalization calculations and is reasonably well
described by a generalized Heitler-London (GHL) approach.
The calculations suggest that electronic correlations
are significant at low magnetic field, leading to spatially
separated single-particle GHL orbitals, reminiscent
of the formation of a Wigner molecule.

The dot is fabricated with electron-beam lithography
defined top-gates on a 55 nm wide parabolic Ga[Al]As
quantum well. The aluminium concentration varies in
growth direction and was chosen such that the Zeemansplitting
of electronic levels is negligible. The particular
gate geometry of the dot (see Fig.\ 1, inset) leads to a
moderate spatial anisotropy of the confinement potential.
Further details about the sample can be found in Ref.\ \cite{yann06}.

Measurements were performed in a dilution refrigerator
with an electronic temperature of 300 mK as determined
from the width of conductance resonances in the
Coulomb blockade regime. Finite bias $I(V)$ traces were
recorded using standard DC current measurement techniques.
Later on the data was numerically differentiated
to obtain the differential conductance $dI/dV_{\rm bias}$.

The sample was tuned into the Coulomb-blockade
regime in the region of the transitions between one and
two, and two and three electrons on the dot. The electron
number was confirmed by measurements of Coulomb-blockade
diamonds \cite{yann06} and by using the on-chip quantum
point contact as a charge detector. In this region, the
dot had a single-particle level spacing of 5 meV and a
charging energy of 6.9 meV.

The magnetic field dependent excitation spectrum
shown in Fig.\ 1 is measured employing tunneling spectroscopy
with varying plunger gate voltage $V_{\rm pg}$ at a fixed
bias voltage of 2.5 mV. We plot the derivative $dI/dV+{\rm pg}$.
Resonances in this quantity correspond to resonances in
the differential conductance $dI/dV_{\rm bias}$. Two families of
resonances can be seen corresponding to transitions between
electron numbers $N = 1$ and 2, and $N = 2$ and 3.

First we concentrate on the $N = 1$ and 2 region. At
zero magnetic field we observe -- in addition to the wellknown
singlet ground state $S_0$ and triplet excited state
$T_+$ that can be seen in Fig.\ 1 -- an additional triplet state
$T_-$ split from the $T_+$ state by the confinement anisotropy
(not seen in the figure, see \cite{yann06}). At finite magnetic fields
we find an additional excited singlet state $S_2$ (see Fig.\ 1)
which shows an avoided crossing with $S_0$ at a magnetic
field beyond the singlet-triplet transition (at 4 T) in the
ground state of the system. Another excited state $T_{+,{\rm CM}}$
even higher in energy than $S_2$ is found and attributed to
a triplet state combining an excitation in the relative and
the center of mass motion (see below). In the transition
region between $N = 2$ and 3 we observe resonances that
correspond to transitions between the two-electron states
$S_0$ and $T_+$ and the three-electron ground state.

%------------------------------ begin figure 1 ------------------------
\begin{figure}[t]
\centering\includegraphics[width=8.cm]{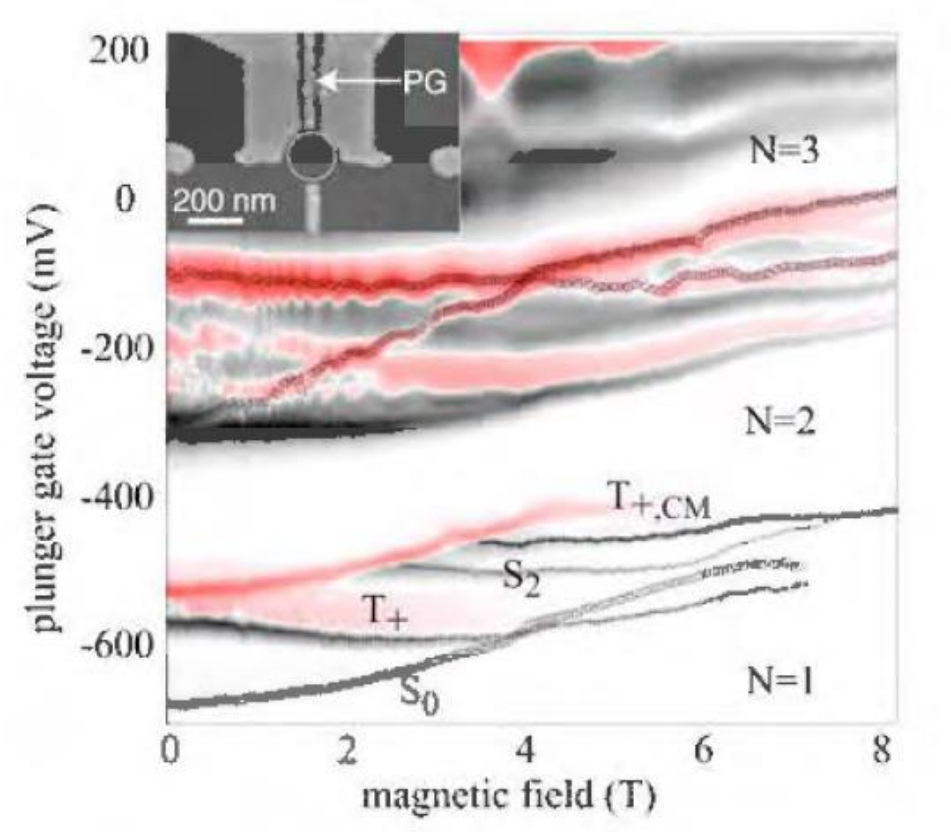}
\caption{%%%%
Measured excited state spectra as a function of magnetic field 
and plunger gate voltage. Inset: sample geometry.
}
\label{fig1}
\end{figure}
%------------------------------ end figure 1 ------------------------

%------------------------------ begin figure 2 ------------------------
\begin{figure}[t]
\centering\includegraphics[width=8.cm]{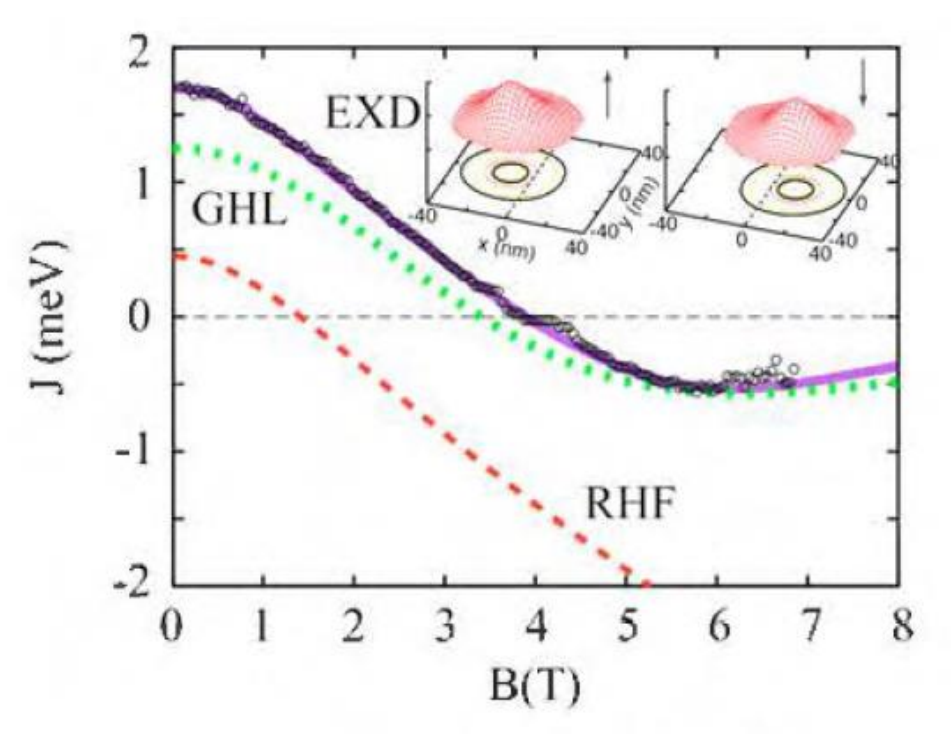}
\caption{%%%%
Comparison of $J(B)$ calculated with different methods and the experimental 
results. Inset: single-particle orbitals (modulus square) of the GHL approach.
}
\label{fig2}
\end{figure}
%------------------------------ end figure 2 ------------------------

The experimental findings can be quantitatively interpreted
by comparing to the results of exact diagonalization
(EXD) calculations for two electrons in an
anisotropic harmonic confinement potential. Details of
the model can be found in Ref.\ \cite{yann06}. The calculated magnetic
field dependent energy splitting $J(B) = E_{T_+} - E_{S_0}$
between the two lowest states $S_0$ and $T_+$ is found to be in
remarkable agreement with the experiment. All the additional
excited states observed in the experiment can be
unambiguously identified with calculated excited states
of the two-electron dot \cite{yann06}.

The EXD calculations give strong evidence for the
importance of correlation effects. Closer inspection of
the total electron densities and conditional probabilities
(CPDs) reveals a strongly correlated ground state at zero
magnetic field, in which the elecu·ons do not occupy
the same single-particle state. Already at magnetic fields
below the singlet-triplet transition, the CPDs indicate
localization of the two electrons and formation of a state
resembling an H$_2$-like Wigner molecule.

We gain further insight into the importance of correlations
in the system by comparing the results for $J(B)$ calculated
exactly and within different approximations. Figure
2 shows the predictions obtained from EXD calculations,
restricted Hartree-Fock (RHF) calculations and
from a generalized Heitler-London (GHL) approach in
comparison with the measured data. Details about the approximative
calculations can be found in Ref.\ \cite{yann990202}. The RHF
and GHL schemes have the advantage that they minimize
the energy using single-particle orbitals. It is evident
from Fig.\ 2 that the RHF method, which assumes
that both electrons occupy the same single-particle orbital,
is not able to reproduce the experimental findings.
The GHL approach, which allows the two electrons to
occupy two spatially separated states, appears to be a
good approximation. Plotting the two single-particle orbitals
resulting from this approach clearly demonstrates
that the two electrons do not occupy the same spatial orbital,
but rather fill single-particle states that are spatially
separated significantly (inset).

\begin{acknowledgements}
We thank R. Nazmitdinov for valuable discussions. Support
from the Schweizerischer Nationalfonds, the US
D.O.E. (Grant No. FG05-86ER45234), and the NSF
(Grant No. DMR-0205328) is gratefully acknowledged.
\end{acknowledgements}

\begin{widetext}

\noindent
\textcolor{red}{THE PUBLISHED VERSION CAN BE DOWNLOADED FROM:}\\
\textcolor{blue}{\underline%%%%%%%%%%%%
{https://cpn-us-w2.wpmucdn.com/sites.gatech.edu/dist/8/1647/files/2021/02/Ihn\_2007\_aipconf.pdf} }\\
~~~~~~\\
\noindent
Cite as: AIP Conference Proceedings {\bf 893}, 783 (2007); 
\textcolor{blue}{\underline{https://doi.org/10.1063/1.2730124}}\\
Published Online: 04 May 2007
\end{widetext}

\end{document}